\documentclass[a4paper,11pt]{article}
\pdfoutput=1 

\usepackage{jinstpub}

\usepackage{lineno}
\usepackage{graphicx}
\usepackage{subcaption}
\usepackage{booktabs} 
\usepackage[export]{adjustbox}
\usepackage{array}
\newcolumntype{P}[1]{>{\centering\arraybackslash}p{#1}}
\usepackage{floatrow}
\graphicspath{ {./images/} }

\usepackage[utf8]{inputenc}
\usepackage{siunitx}
\usepackage{gensymb}
\usepackage{wasysym}
\usepackage{upgreek}

\title{Performance of Prototypes with Different Reflector Materials for the SHiP Liquid Scintillator Surrounding Background Tagger} 

\author[a,*]{A.~Brignoli}
\author[d]{, P.~Deucher}
\author[a]{, C.~Eckardt}
\author[b]{, F.~Faller}
\author[b]{, H.~Fischer}
\author[d,*]{, A.~Hollnagel}
\author[b]{, A.~Krolla}
\author[a,*]{, H.~Lacker}
\author[b,*]{, F.~Lyons}
\author[d]{, J.~Molins i Bertram}
\author[b]{, T.~Molzberger}
\author[b]{, A.~S.~Müller}
\author[b]{, S.~Ochoa}
\author[a]{, A.~Reghunath}
\author[b]{, T.~Rock}
\author[c]{, M.~Schaaf}
\author[a]{, C.~Scharf}
\author[b]{, M.~Schumann}
\author[b]{, J.~M.~Webb}
\author[b]{, J.~Wenk}
\author[a]{, I.~Wöstheinrich}
\author[d]{, M.~Wurm}

\emailAdd{brignoli@physik.hu-berlin.de}
\emailAdd{annika.hollnagel@uni-mainz.de}
\emailAdd{lacker@physik.hu-berlin.de}
\emailAdd{fairhurst.lyons@physik.uni-freiburg.de}

\affiliation[a]{Institut für Physik, Humboldt-Universität zu Berlin, Newtonstraße 15, 12489 Berlin, Germany}
\affiliation[b]{Physikalisches Institut, Albert-Ludwigs-Universität Freiburg, Hermann-Herder-Straße 3, 79104 Freiburg, Germany}
\affiliation[c]{Zentralinstitut für Engineering, Elektronik und Analytik -- Engineering und Technologie, Forschungszentrum Jülich GmbH, Wilhelm-Johnen-Straße, 52428 Jülich, Germany}
\affiliation[d]{Institut für Physik \& Exzellenzcluster PRISMA$^+$, Johannes Gutenberg-Universität Mainz, Staudingerweg 7, 55128 Mainz, Germany}
\affiliation[*]{Corresponding authors}

\date{March 2025}

\abstract{The baseline technology for the Surrounding Background Tagger of the recently approved SHiP experiment relies on liquid scintillator composed of linear alkylbenzene and \hbox{2,5-diphenyloxazole} as active detector material. The primary scintillation photons are collected by Wavelength-shifting Optical Modules, and the secondary photons are guided by total reflection to an array of 40~silicon photomultipliers.

Here we present a direct comparison of the performance of three detector prototype cells constructed from different alternative materials known to provide good chemical compatibility with the liquid scintillator: Aluminium and stainless steel. To increase the internal reflectivity, one of the two aluminium prototypes was polished on the inside, while the inner walls of the stainless steel cell were clad with sheets of polytetrafluoroethylene. Using 5\,GeV muons from the CERN~PS~T10 beamline, we studied detected light yield and time resolution attained by the three prototypes. For both the polished AlMg4.5 cell and the PTFE-clad stainless steel prototype, the achieved detected light yield and time resolution meet the requirements of the SHiP Surrounding Background Tagger. Concerning another crucial parameter, the uniformity of the detector response across the detector cell, the polished AlMg4.5 cell shows the best performance among the tested prototypes. These results will significantly affect the final design of the SHiP Surrounding Background Tagger.}

\begin{document}
\maketitle

\section{Introduction}\label{Sec:Introduction}

The SHiP (Search for Hidden Particles) experiment was recently approved by CERN to be built at a future Beam Dump Facility in ECN3 at the SPS North Area~\cite{SHiP:2015vad,SHIP:2021tpn,SHiP-LoI,SHiP-Proposal}. The science goal of SHiP is the production of very rare beyond Standard Model particles from the so-called Hidden Sector - such as heavy neutral leptons and dark photons - and the subsequent detection of their decay signatures in a $\sim$50\,m long Decay Volume equipped with a spectrometer~\cite{Alekhin:2015byh,Bondarenko:2023fex}. The original SHiP proposal planned for the evacuation of the large Hidden Sector Decay Volume for maximum suppression of backgrounds. After detector optimisation and feasibility studies, the new baseline design incorporates a Decay Volume filled with Helium at atmospheric pressure to minimise the interactions of neutrinos inside. 

To achieve the zero background goal of SHiP, a set of veto detectors is required. One of them is the Surrounding Background Tagger (SBT), see Sec.~\ref{sec:sbt}, which detects background particles entering the Decay Volume from the sides via scintillation light. In this work we present test beam results on the performance of detector prototype cells using different light reflector materials that directly impact the detection threshold and thus affect the SBT's veto efficiency.

\subsection{The SHiP Liquid Scintillator-Surrounding Background Tagger}\label{sec:sbt}

The SHiP Hidden Sector Decay Volume will be enveloped by the Surrounding Background Tagger~(SBT).
This allows tagging of muons with energies from 1 -- 400\,GeV entering the detector from the outside, as well as reactions of muons and neutrinos with the gas inside the Decay Volume and its envelope. For optimal hermeticity, the SBT veto detector will use as active detector material a 20\,cm thick layer of liquid scintillator~(LS) composed of the solvent linear alkylbenzene~(LAB) and the fluorescent 2,5-diphenyloxazole~(PPO) at a concentration of 2.0\,g/l. This provides a high interaction probability for particles crossing the SBT, in particular neutral hadrons which are produced in neutrino and muon interactions within the helium-filled Decay Volume. To detect the scintillation light generated in the LS, Wavelength-shifting Optical Modules (WOMs)~\cite{Bastian-Querner:2021uqv} will be employed. Those are made from PMMA tubes coated with a wavelength-shifting dye and coupled to an array of silicon photomultipliers (SiPMs) at one end.

The SBT must provide a very high detection efficiency for the energy depositions of minimum ionising particles (MIPs) crossing the liquid scintillator layer even at smallest path length. As benchmark, background rejection studies of muon and neutrino deep inelastic interactions in SHiP assume a detection efficiency of $>$~99\% for an energy threshold of 45\,MeV. Further requirements for the SBT are good time resolution in the range of 1\,ns and a spatial resolution of a few 10\,cm to reconstruct the origin of background interactions in both time and space~\cite{SHiP-LoI}. The proof-of-principle for the detector technology was demonstrated in \cite{Ehlert:2018pke}, followed by further dedicated test beam campaigns with full-scale detector prototypes at the DESY~II test beam facility in 2022~\cite{Alt:2023vuu} and later at CERN, confirming the detector performance and meeting these benchmarks.

\subsection{Preceding Studies}

Following the initial design of the SHiP Decay Volume, previous LS-SBT prototypes were constructed from sheets of S355JO(J2/K2)W Corten steel, with a cell size of about 120 $\times$ 80 $\times$ 20\,cm$^3$ and instrumentation with two WOM tubes~\cite{Alt:2023vuu}. To reach the required high values of detected light yield, the inner walls of the detector cells had to be spray-coated through the two WOM openings, first with a primer, followed by a highly reflective paint containing Barium Sulfate (BaSO$_4$). When concluding the test beam exposure of the first prototype~\cite{Alt:2023vuu}, a significant amount of rust staining was observed on the inner surface of the coated Corten steel walls. Subsequent tests with a small laboratory mock-up confirmed that this was caused by increased drying time of the water-based acrylic primer inside the virtually closed volume of the cell. Consequently, a non-acrylic anti-rust primer was applied to the inner walls of the next detector prototype. Although these measures significantly improved the surface quality, still some rust stains appeared that directly affected the cell wall reflectivity and also led to significant ageing of the LS. Moreover, spray-coating of the inner cell walls through the WOM openings is a non-trivial task and costly both in terms of spray-coating supplies and time. Consequently, the new work presented here investigates alternative materials that do not require coating.

\section{Detector Prototypes}

\begin{figure}[ht]
    \centering
    \begin{subfigure}[c]{0.49\textwidth}
        \centering
        \includegraphics[height=8.2cm]{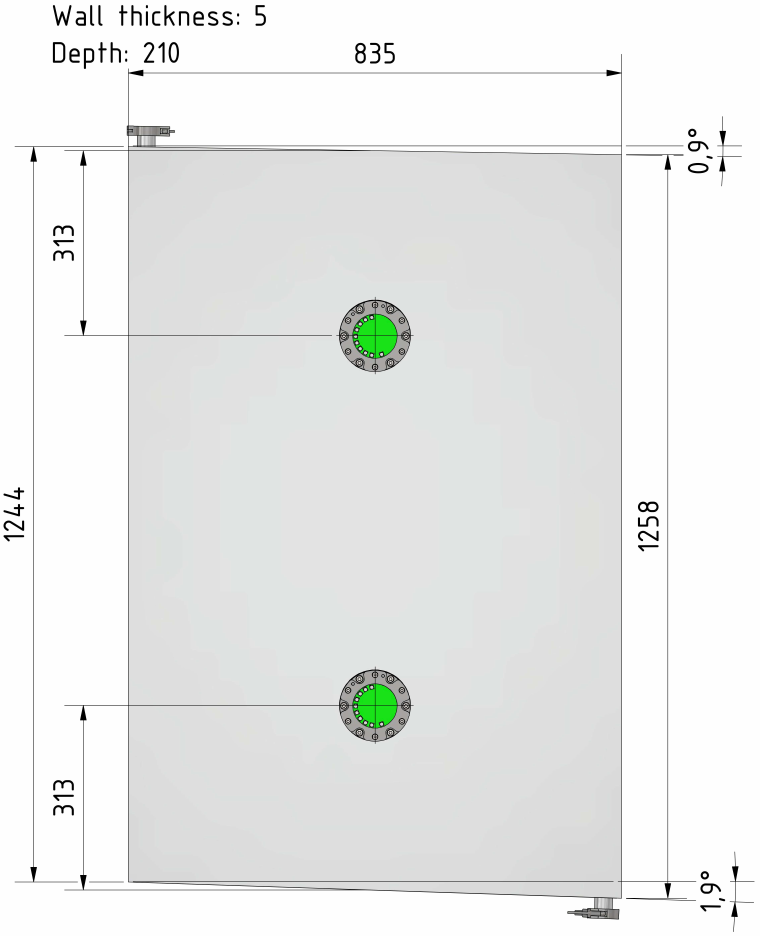}
        \caption{}\label{fig:CellDimensionsAandC}
    \end{subfigure}
    \begin{subfigure}[c]{0.49\textwidth}
        \centering
        \includegraphics[height=8.2cm]{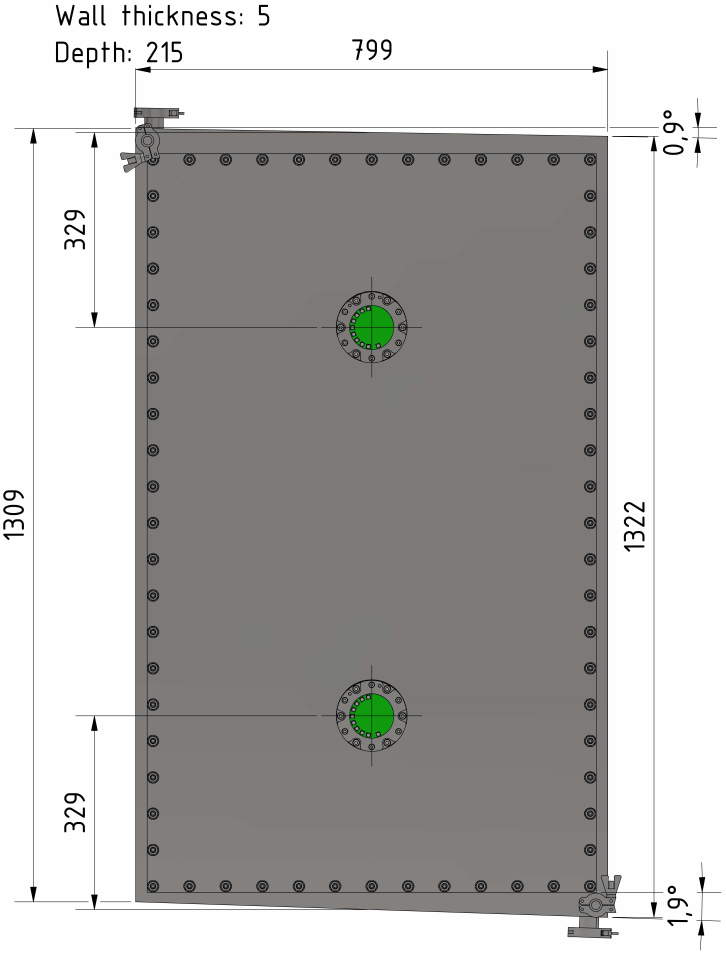}
        \caption{}\label{fig:CellDimensionsB}
    \end{subfigure}
    \caption{Technical drawings of the prototype cells. \textit{Left}: Cells A (polished AlMg4.5) and C (unpolished AlMg4.5). Cell~A was manually polished on the inside before the lid was welded on. \hbox{\textit{Right}: Cell~B} (stainless steel + PTFE). To allow for post welding installation of the internal PTFE reflector sheets, the lid housing the light detectors was screwed on, sealed by a large O-ring of Viton\textsuperscript{\texttrademark}.}\label{fig:CellDimensions}
\end{figure}

Three new LS-SBT prototype cells with dimensions of around 120 $\times$ 80 $\times$ 20\,cm$^3$ were built using different materials and surface finishes. Following the frustrum shape of the Decay Volume, these cells are not perfectly rectangular: the long vertical sides are parallel with a slight shift, and there is a small angle of 2.8$^\circ$ between the shorter sides (see Fig.~\ref{fig:CellDimensions}).

\subsection{Material Specifications}\label{sec:MaterialSpecs}

\begin{figure}[ht]
    \begin{subfigure}[c]{0.49\textwidth}
        \centering
        \includegraphics[width=1.0\textwidth]{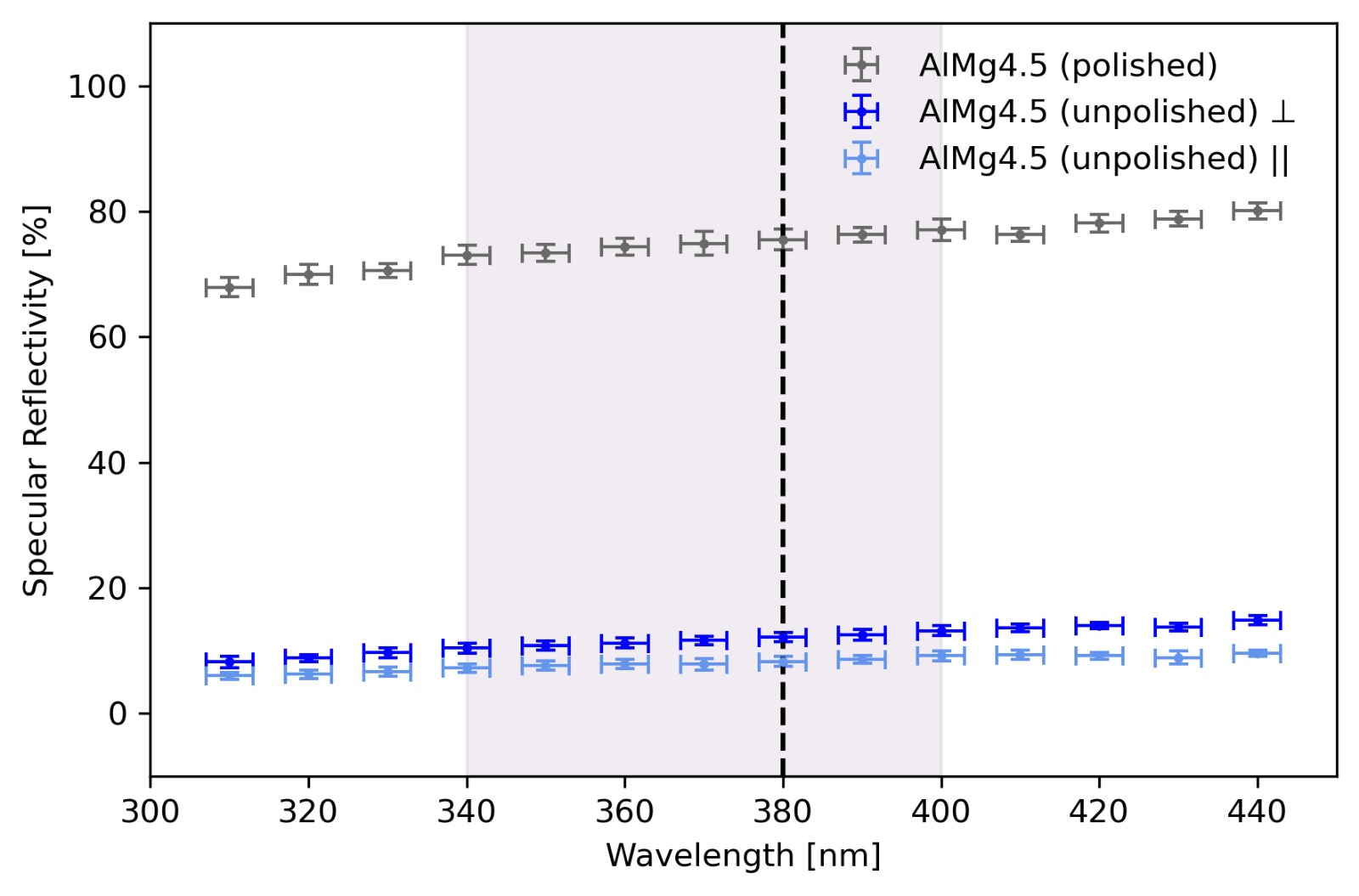}
       \caption{Specular reflector materials.}\label{fig:reflectivity-specular}
    \end{subfigure}
    \hspace{0.1cm}
    \begin{subfigure}[c]{0.49\textwidth}
        \centering
        \includegraphics[width=1.0\textwidth]{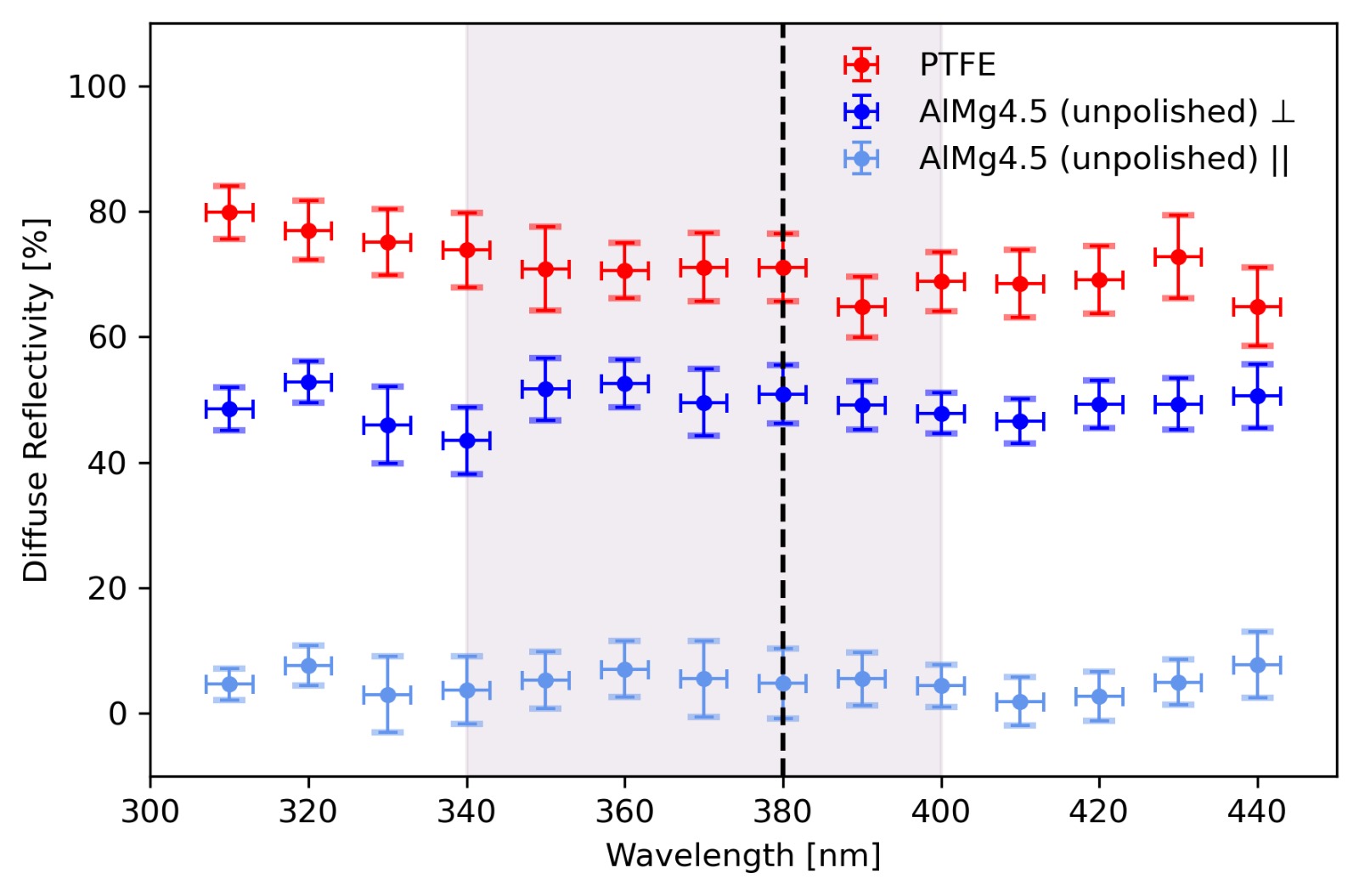}
        \caption{Diffuse reflector materials.}\label{fig:reflectivity-diffuse}
    \end{subfigure}
    \caption{Laboratory measurements of specular (\textit{left}) and diffuse (\textit{right}) reflectivity of the materials used for the detector prototype cells. The specular reflector polished AlMg4.5 [\textit{grey}] corresponds to the inner surface quality of Cell~A, exhibiting a reflectivity of (75.2~$\pm$~1.6)\% at the benchmark wavelength of 380\,nm. The 0.5\,mm thick sheet of PTFE [\textit{red}] used in Cell~B shows a diffuse reflectivity of (71.0~$\pm$~5.4)\% at 380\,nm. Raw unpolished AlMg4.5 [\textit{blue}], used in Cell~C, exhibits neither pure specular nor pure diffuse reflectivity due to the manufacturing process of rolling. This results in a strong anisotropy depending on the angle $\theta$ of incidence plane w.r.t.~the furrowed surface structure: The measured specular reflectivity at 380\,nm is (12.2~$\pm$~0.7)\% and (8.3~$\pm$~0.8)\% for $\theta=90\degree$ and $\theta=0\degree$, respectively, while the diffuse reflectivity is (50.9~$\pm$~4.6)\% and (4.8~$\pm$~5.6)\%. The shaded region at 340 -- 400\,nm indicates the emission range of the LS fluorescent PPO convoluted with the absorption spectrum of the WOM wavelength-shifter, with the black dashed line highlighting the benchmark wavelength of 380\,nm.}
   \label{fig:reflectivity}
\end{figure}

As alternatives to the previously used Corten steel base material, the following materials were studied:
\begin{itemize}
    \item \textbf{Aluminium (AlMg4.5):} This widely used alloy of aluminium is chemically compatible with the LS and naturally exhibits the highest UV reflectivity of any metal. However, from simulations it was expected that the reflectivity of untreated rolled AlMg4.5 would still be insufficient for meeting the SBT performance requirements. It was thus decided to construct two detector cells from AlMg4.5: one from raw rolled material (hereafter referred to as "unpolished") and one from rolled material with manually polished inner walls (hereafter referred to as "polished") providing a very high specular reflectivity.
    \item \textbf{Stainless steel + PTFE reflector:} Another material known to have very good chemical compatibility with the LS is stainless steel. However, due to its poor reflectivity in the relevant wavelength range of 340 -- 400\,nm, photon transport simulations showed that even a polished surface would result in insufficient light collection. Consequently, a new detector prototype was manufactured from stainless steel, featuring a large rectangular flange with dimensions 760 $\times$ 1\,250\,mm$^2$. Using screws, 0.5\,mm thick PTFE sheets were mechanically fastened to the inner walls of this detector to provide a highly diffusive-reflective lining. PTFE was chosen due to its excellent reflectivity in the ultraviolet range, which is critical for maximising the light collection efficiency.
\end{itemize}
With a dedicated laboratory setup, the specular and diffuse reflectivities of several materials were studied, including unpolished and manually polished samples of AlMg4.5 and 0.5\,mm thick sheets of PTFE~\cite{ptfe} on stainless steel, as used in the detector prototypes (see Fig.~\ref{fig:inside-cells}). The results of these measurements are shown in Fig.~\ref{fig:reflectivity}: at the benchmark wavelength of 380\,nm - determined by the spectral characteristics of the combined LS~+~WOM system - the specular reflectivity of polished AlMg4.5 is measured as 75\%, while the diffuse reflectivity of the 0.5\,mm thick PTFE sheet is around~71\%. The reflectivity of unpolished AlMg4.5 is highly angular dependent and considerably lower.

\subsection{General Prototype Setup}

\begin{figure}[htb]
    \begin{subfigure}[c]{0.305\textwidth}
        \centering
        \includegraphics[width=1.0\textwidth]{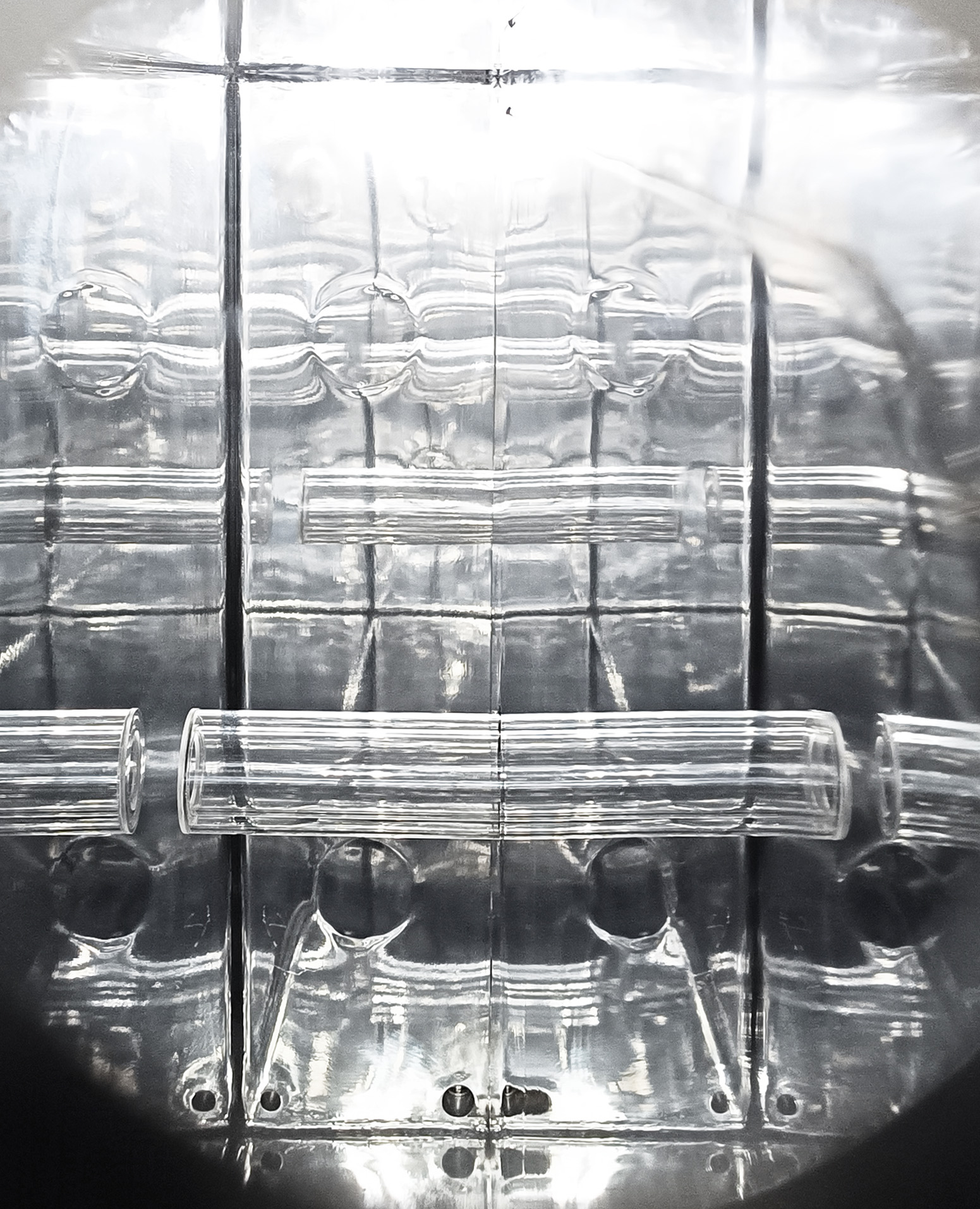}
        \caption{}\label{fig:inside-cell-a}
    \end{subfigure}
    \hspace{0.4cm}
    \begin{subfigure}[c]{0.305\textwidth}
        \centering
        \includegraphics[width=1.0\textwidth]{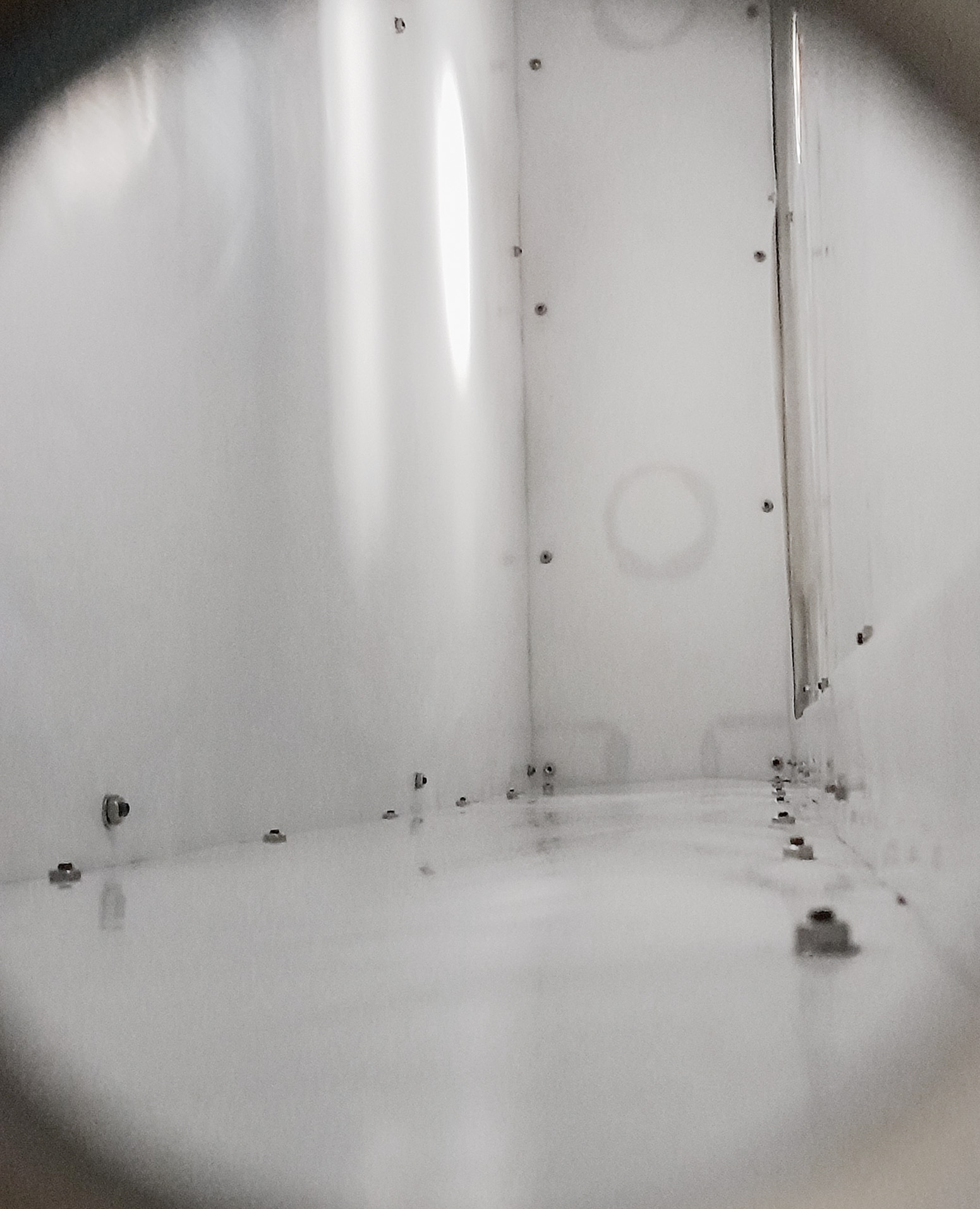}
        \caption{}\label{fig:inside-cell-b}
    \end{subfigure}
    \hspace{0.4cm}
    \begin{subfigure}[c]{0.305\textwidth}
        \centering
        \includegraphics[width=1.0\textwidth]{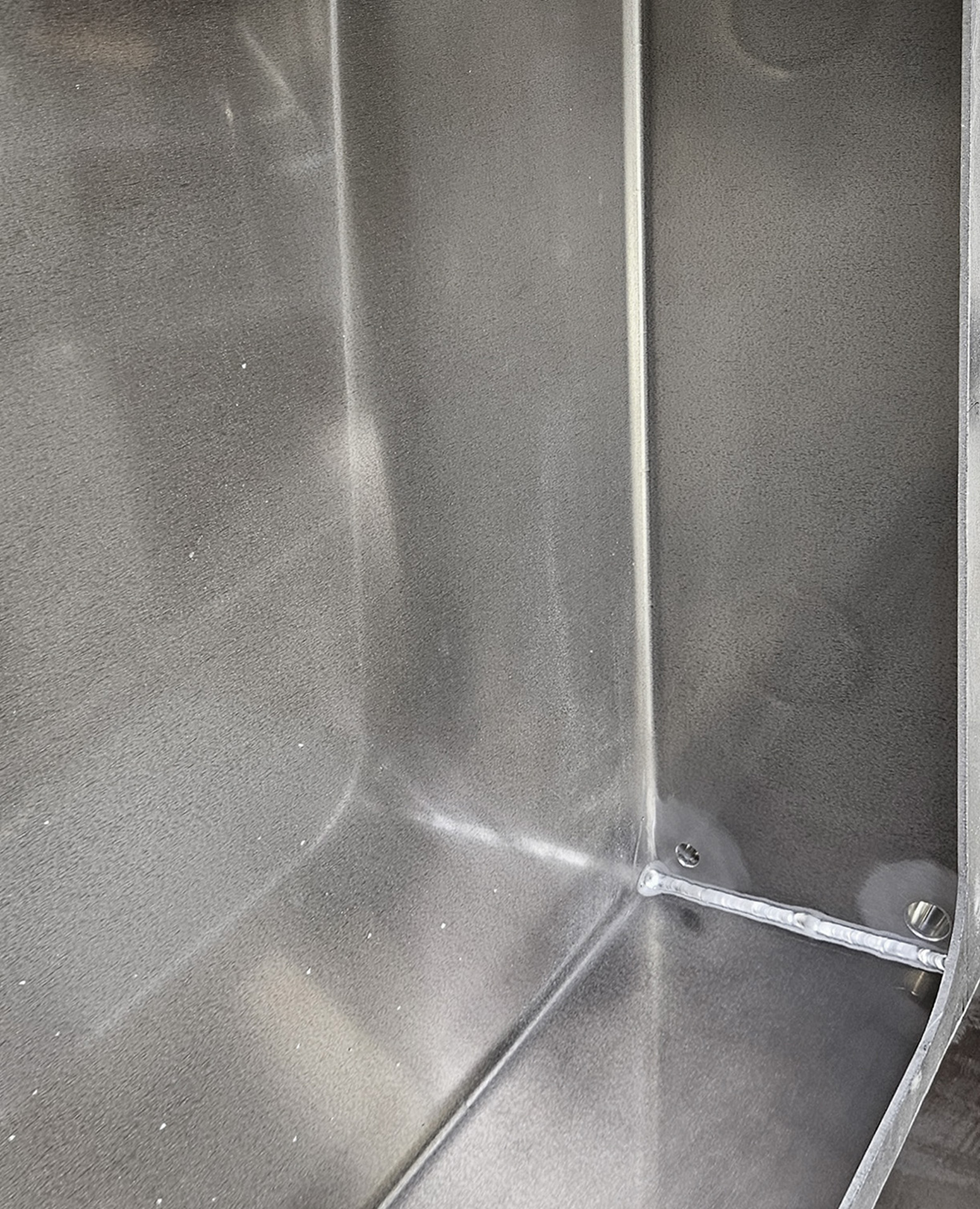}
        \caption{}\label{fig:inside-cell-c}
    \end{subfigure}
    \caption{Inside view of the empty detectors before filling with LS. \textit{Left}: Surface of Cell~A (polished AlMg4.5) after welding. Multiple reflections demonstrate the high specular reflectivity. \textit{Middle}: Surface of Cell~B (stainless steel, clad with 0.5\,mm thick sheets of PTFE). \textit{Righ}t: Surface of Cell~C (unpolished AlMg4.5) before welding the final plate. The material exhibits a strongly anisotropic diffuse reflectivity.}\label{fig:inside-cells}
\end{figure}

In the following, we will denote the detector cell made from polished AlMg4.5 as "Cell~A", the prototype made from stainless steel + PTFE  "Cell~B", and the detector constructed from unpolished AlMg4.5 "Cell~C". All three cells were of comparable geometry and dimensions as given in Fig.~\ref {fig:CellDimensions}. The internal surfaces are shown in Fig.~\ref{fig:inside-cells}.

\begin{figure}[ht]
    \centering
    \includegraphics[width=0.65\textwidth]{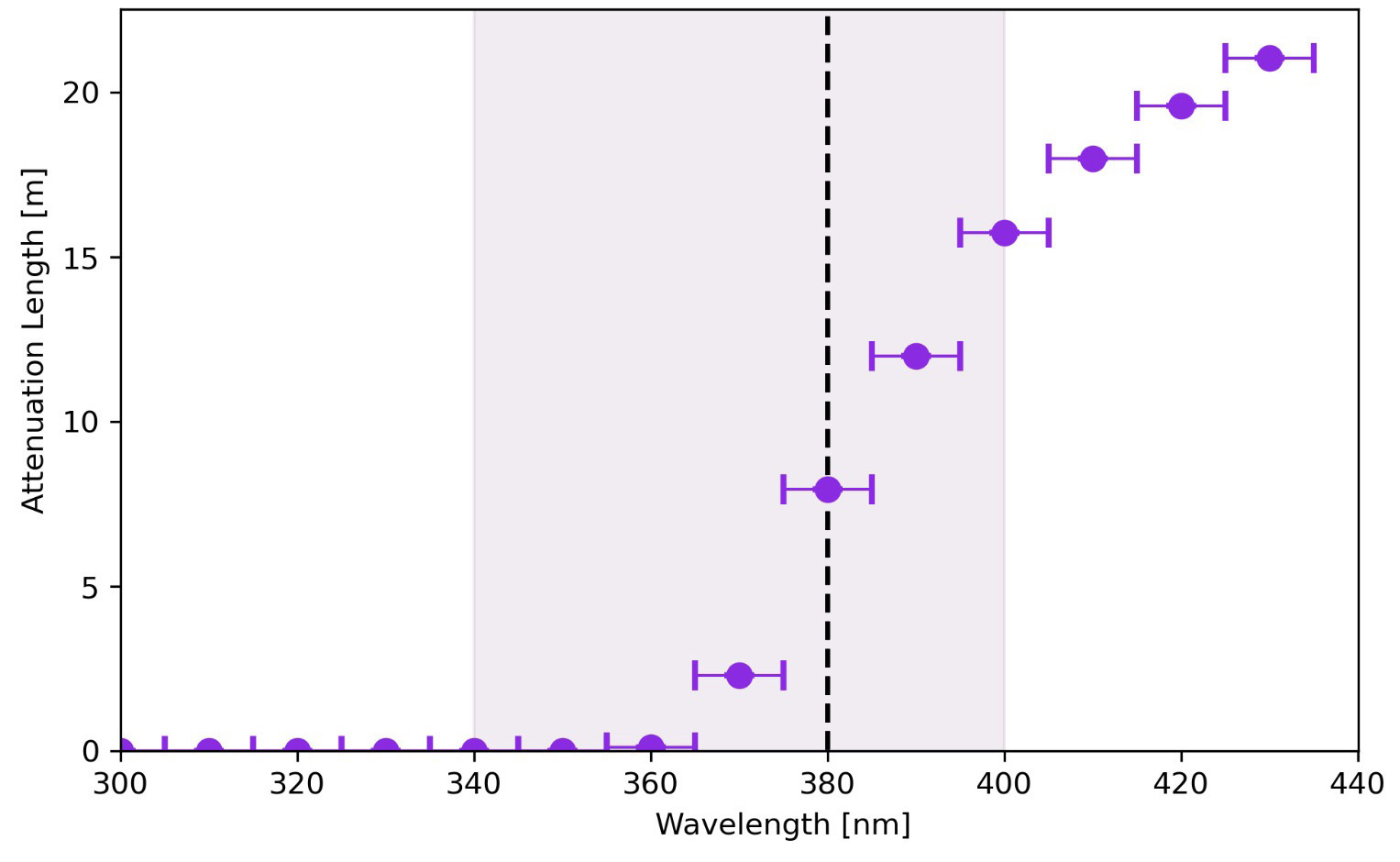}
    \caption{Measurement of the attenuation length of the liquid scintillator in the detector prototypes that was also used as input for the detector cell simulation. At 380\,nm, the attained attenuation length is (7.9 $\pm$ 1.5)\,m. To compensate for increasing statistical fluctuations at attenuation lengths~$>$15\,m, the measurements have been averaged over a range of 10\,nm, as indicated by the horizontal bars on the data points.}
   \label{fig:tb2024_ls_absorption}
\end{figure}

The experimental setups were similar to those used for the previous prototype~\cite{Alt:2023vuu} constructed from spray-coated Corten steel:
\begin{itemize}
    \item Each prototype cell was filled with LS made from alumina-purified LAB plus 2.0\,g/l of PPO, with butylated hydroxytoluene (BHT) added as anti-oxidant at a concentration of 0.015\%. The absorbance of the LS was measured with a PerkinElmer $\Lambda$850 spectrophotometer and a Suprasil glass cuvette of 100\,mm path length, then translated into attenuation length as de\-scribed in~\cite{Alt:2023vuu}. At 380\,nm, an attenuation length of (7.9 $\pm$ 1.5)\,m could be attained (see~Fig.~\ref{fig:tb2024_ls_absorption}).
    \item Each detector cell was instrumented with two WOMs made from PMMA tubes with a length of 205\,mm, an outer diameter of 60\,mm, and a wall thickness of 3\,mm. Each WOM tube was dip-coated on its outside and inside with a dye made from toluene and Paraloid~B723~(PEMA) as base material, and the two wavelength shifters bis-MSB and \hbox{p-terphenyl (PTP)}. The WOMs were housed in PMMA vessels and fixed to the respective flanges on the detector cells (see~Fig.~\ref{fig:CellDimensions}) with a structure optimized to reduce light losses.
    \item Each of the WOMs was instrumented at the flange end with a ring-shaped array of 40 SiPMs of 3\,mm~$\times$~3\,mm (HPK S14160-3050HS), operated at an overvoltage of about 3.7\,V.
    \item Each ring-shaped SiPM board was connected to a PCB hosting an eMUSIC chip~\cite{MUSIC-ASIC} for preamplification of the SiPM signals which were combined to 8~groups of 5~SiPMs per WOM. The waveforms of these signals were recorded with a WaveCatcher digitiser~\cite{WaveCatcher}.
\end{itemize}
\smallskip

\section{Test Beam Exposure of the Detector Prototypes}\label{Sec:CERNtestbeam}

The three cells were tested in November 2024 at the CERN PS T10 beamline with a 5\,GeV muon beam delivering MIPs. 

\subsection{Measurement Setup}\label{Sec:DetectorSetup} 

\begin{figure}[ht]
    \begin{subfigure}[c]{0.64\textwidth}
        \centering
        \includegraphics[width=1.0\textwidth]{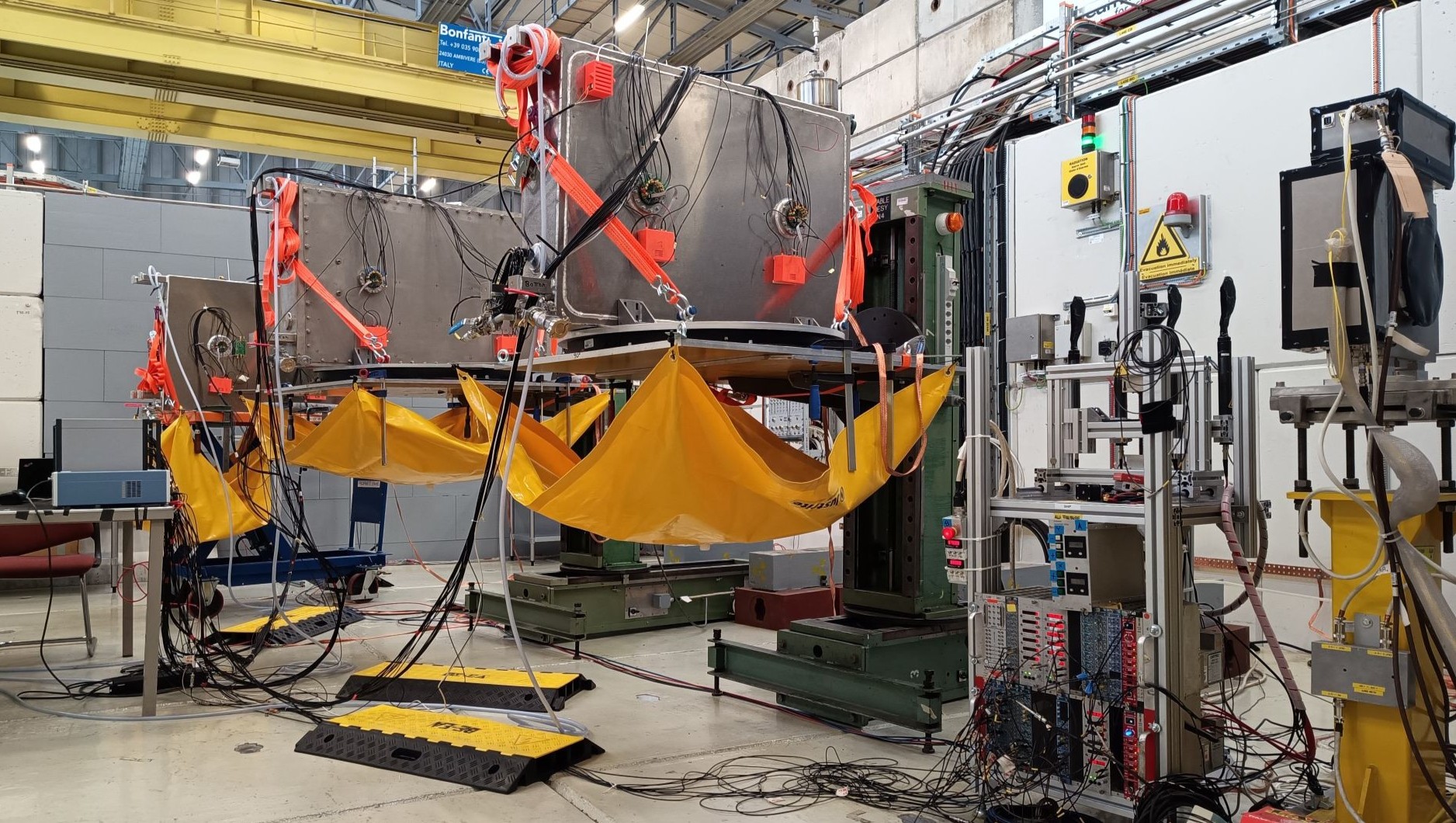}
    \end{subfigure}
    \hspace{0.3cm}
    \begin{subfigure}[c]{0.325\textwidth}
        \centering
        \includegraphics[width=1.0\textwidth]{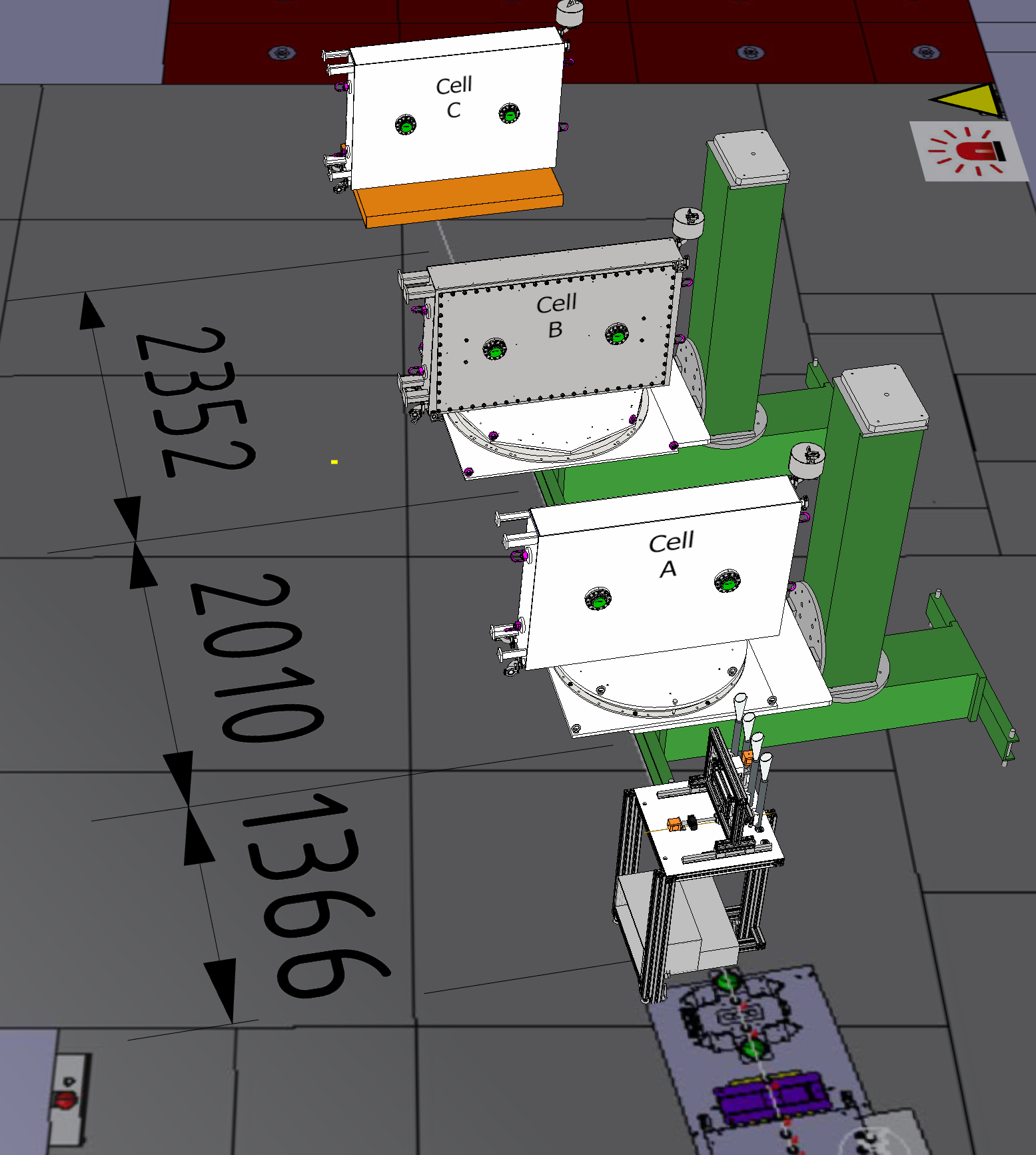}
    \end{subfigure}
    \caption{Prototype Cells~A, B, and C installed in the CERN PS T10 experimental area. Actual detector setup during the measurements (\textit{left}) and technical drawing showing the relative detector locations in mm (\textit{right}). The beam telescope providing the trigger signal of incoming beam particles was located upstream of Cell~A, with an additional plastic scintillator panel mounted downstream of Cell~C (not visible here). The yellow sheets below the cells are part of the LS retention safety~system.}\label{fig:tb-setup}
\end{figure}

The three detector cells were arranged in a line as shown in Fig.~\ref{fig:tb-setup} and positioned such that the particle beam crossed the three detectors at approximately the same point. Cell~A was positioned furthest upstream, followed by Cell~B and Cell~C. In front of the three detector cells, a beam telescope made of four thin plastic scintillators of 5\,cm diameter was placed along the beamline. Between the inner two of these four scintillators an additional detector was placed, made out of two thin plastic scintillators of 1.8\,cm diameter. Behind Cell~C, another 10\,mm thick plastic scintillator of $28 \times 28$\,cm$^2$ area was placed to identify traversing beam particles.
The data acquisition of the WaveCatcher digitiser was triggered by a coincidence of three out of the four telescope scintillators, the two small scintillators inside the beam telescope and the large plastic scintillator behind the setup.

\subsection{Collected Data}\label{Sec:CollectedData}

\begin{figure}[ht]
    \begin{subfigure}[c]{0.49\textwidth}
        \centering
        \includegraphics[width=1.\textwidth]{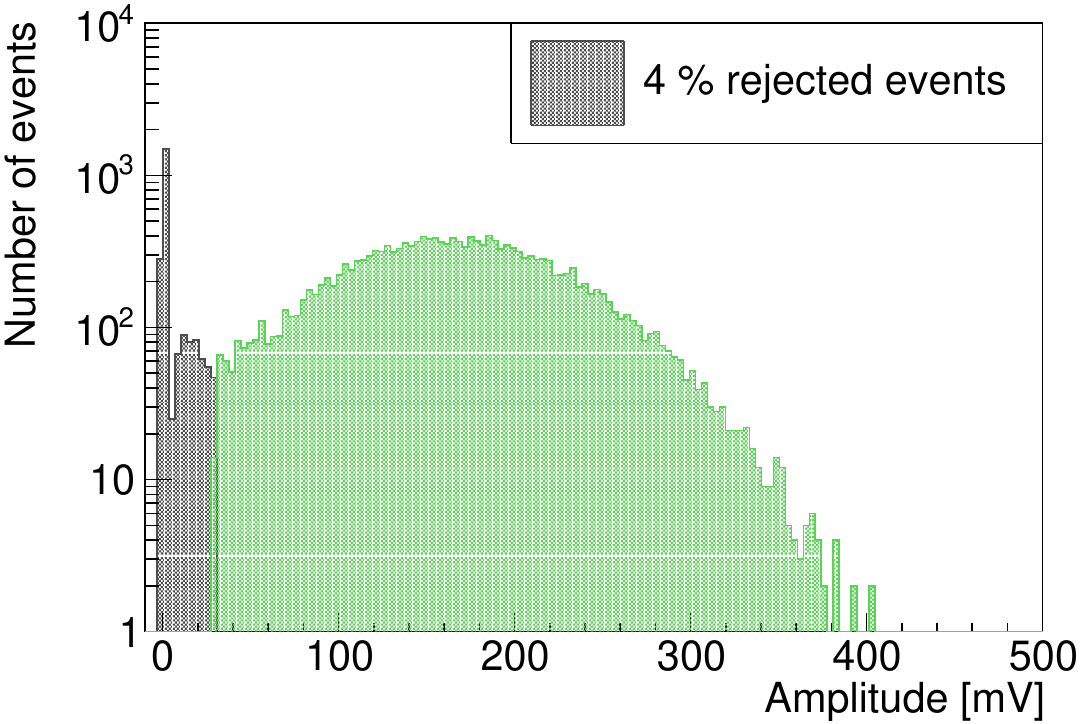}
        \caption{CO$_2$ pressure of 1.1 \,bar.}\label{fig:AmplitudeSpectrumCC1.1bar}
    \end{subfigure}
    \hspace{0.1cm}
    \begin{subfigure}[c]{0.49\textwidth}
        \centering
        \includegraphics[width=1.\textwidth]{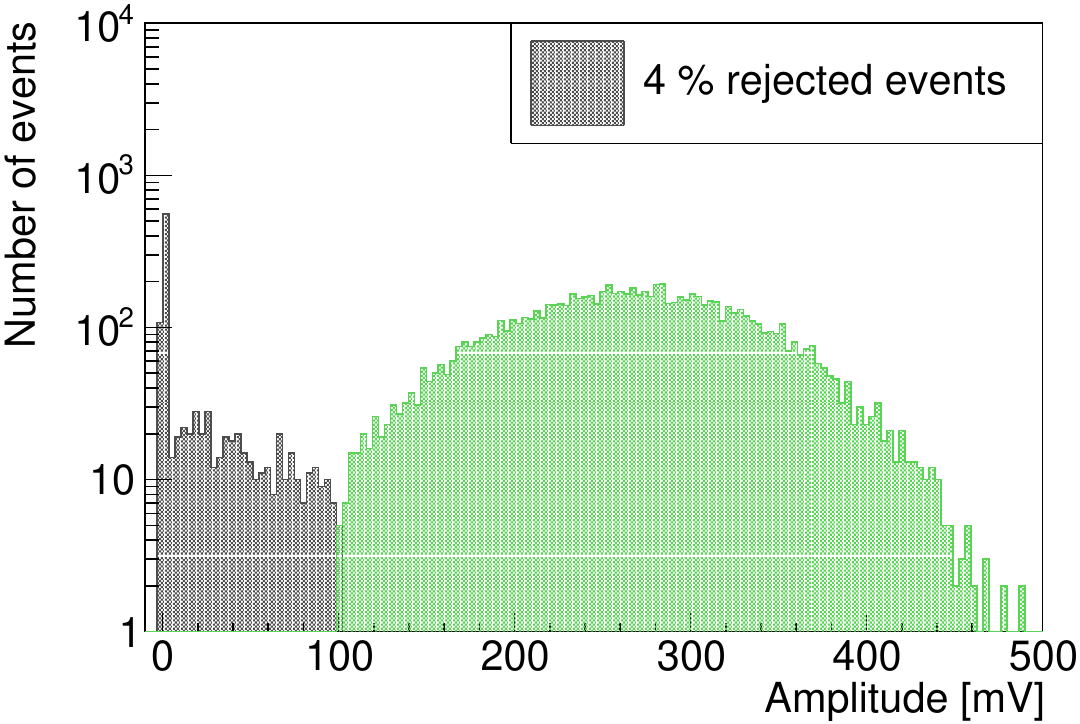}
        \caption{CO$_2$ pressure of 1.5\,bar.}\label{fig:AmplitudeSpectrumCC1.5bar}
    \end{subfigure}
    \caption{Signal amplitude spectrum measured with the Cherenkov Counter for CO$_2$ pressures of 1.1\,bar (\textit{left}) and 1.5\,bar (\textit{right}). The peak at 0\,mV corresponds to a beam contamination with protons and kaons. The region marked \textit{grey} between amplitudes of 0\,mV and 30\,mV for a CO$_2$ pressure of 1.1\,bar (0\,mV and 100\,mV for a CO$_2$ pressure of 1.5\,bar) is assumed to be populated by a contamination of pions in the beam. Events in the \textit{green} region at higher signal amplitudes are expected to originate from muons and constitute about 96\% of the recorded events.}
\label{fig:AmplitudeSpectrumC}
\end{figure}

Measurements were performed with a 5\,GeV muon beam with a closed beam stopper and beam collimator~4 placed in the beamline to reduce the beam's hadron contamination. We also read out the signal from the beam Cherenkov counter (ZT10.XCET040), operated with CO$_2$ at a pressure of 1.1\,bar, which allows rejecting events induced by kaons and protons completely if no Cherenkov signal is observed. At this pressure, however, a sizeable fraction of the pions will still fire the Cherenkov counter. In the offline analysis, one can then separate muon from pion events on a statistical basis, which is important since the pions could initiate hadronic showers along their way through the setup. The corresponding measured Cherenkov counter signal amplitude spectrum is shown in Fig.~\ref{fig:AmplitudeSpectrumCC1.1bar}. By rejecting events with a Cherenkov counter amplitude below 30\,mV, we were able to select a sample of very high muon purity, with a negligible loss of statistics. One run was also performed with a CO$_2$ pressure of 1.5\,bar, which enabled us to estimate the pion, kaon and proton contamination for this beam configuration more precisely than with the CO$_2$ pressure of 1.1\,bar. From the measured Cherenkov counter amplitude spectrum at 1.5\,bar (see Fig.~\ref{fig:AmplitudeSpectrumCC1.5bar}), we estimate a hadron contamination of about 4\% from pions, kaons, and protons for this beam configuration which combined beam collimator~4 and closed beam stopper.

\begin{figure}[ht]
    \centering   
    \includegraphics[width=1.0\textwidth]{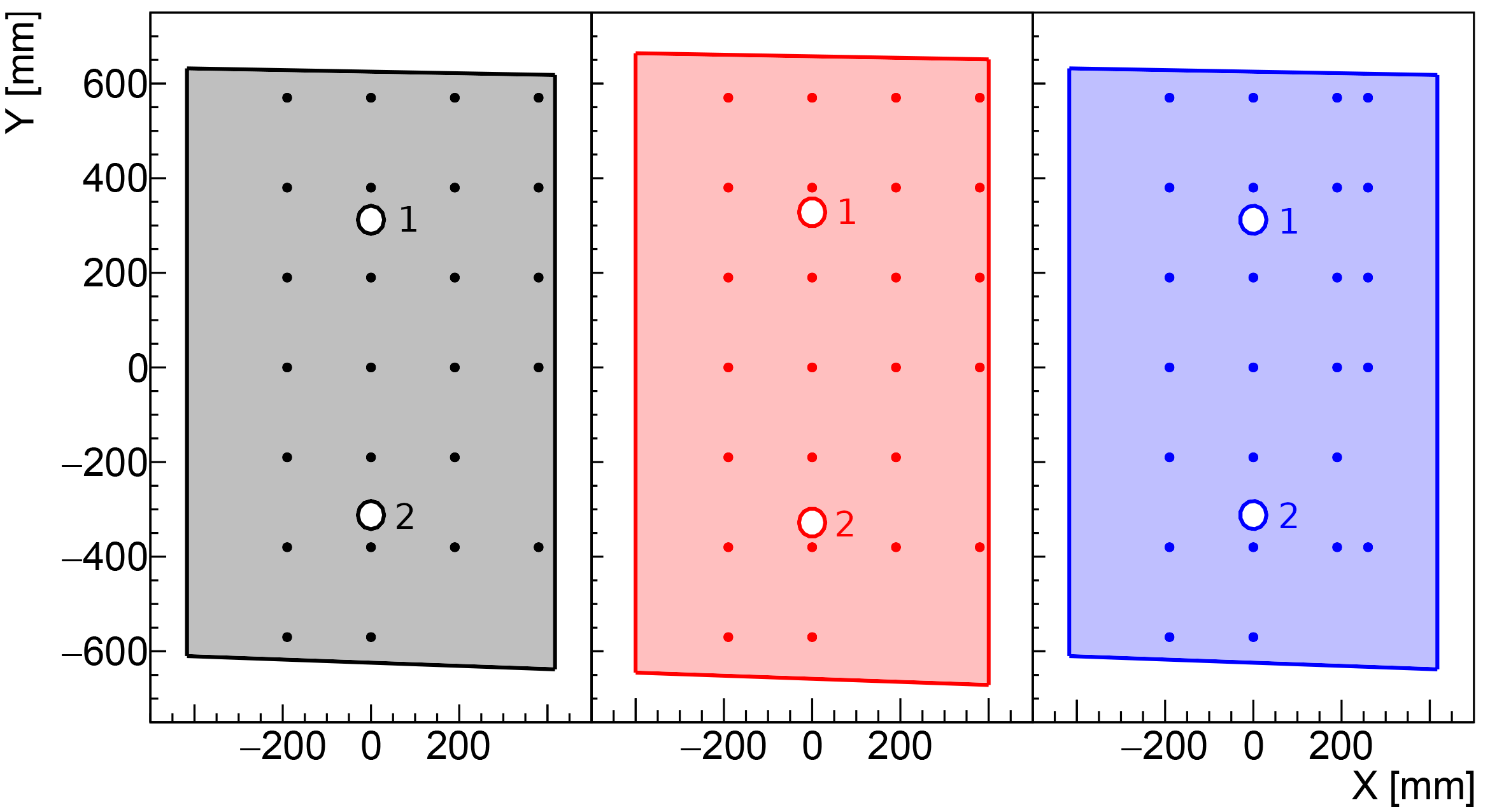}
    \caption{ The three detector cells A (\textit{left}), \hbox{B (\textit{middle)}}, and C (\textit{right}) in their respective coordinate systems together with the measured particle crossing points indicated by the corresponding dots. The WOMs with positive and negative $y$-coordinate are labelled as~1 and~2, respectively.}\label{fig:deglocations}
\end{figure}

Several particle crossing points on the detector cells were measured (indicated in Fig.~\ref{fig:deglocations}), with the beam perpendicular to the cells at an incidence angle of 0$\degree$. For each measurement, 20~000~events were recorded.

\subsection{Monte Carlo Simulation of the Prototype Cells}\label{Sec:MonteCarloSimulation}

A Geant4 simulation~\cite{Agostinelli} of the complete detector cell (using the geometry of Cell~A and C) was developed, taking into account the exact cell dimensions and corresponding wall thicknesses, LS~properties, and relevant parameters for subsequent light transport. More details are described in~\cite{Alt:2023vuu}. Using this simulation, the light collected by the SiPMs on the two WOMs can be predicted for different inner surface reflectivities of the cell walls.

\begin{figure}
    \centering
    \includegraphics[width=0.9\textwidth]{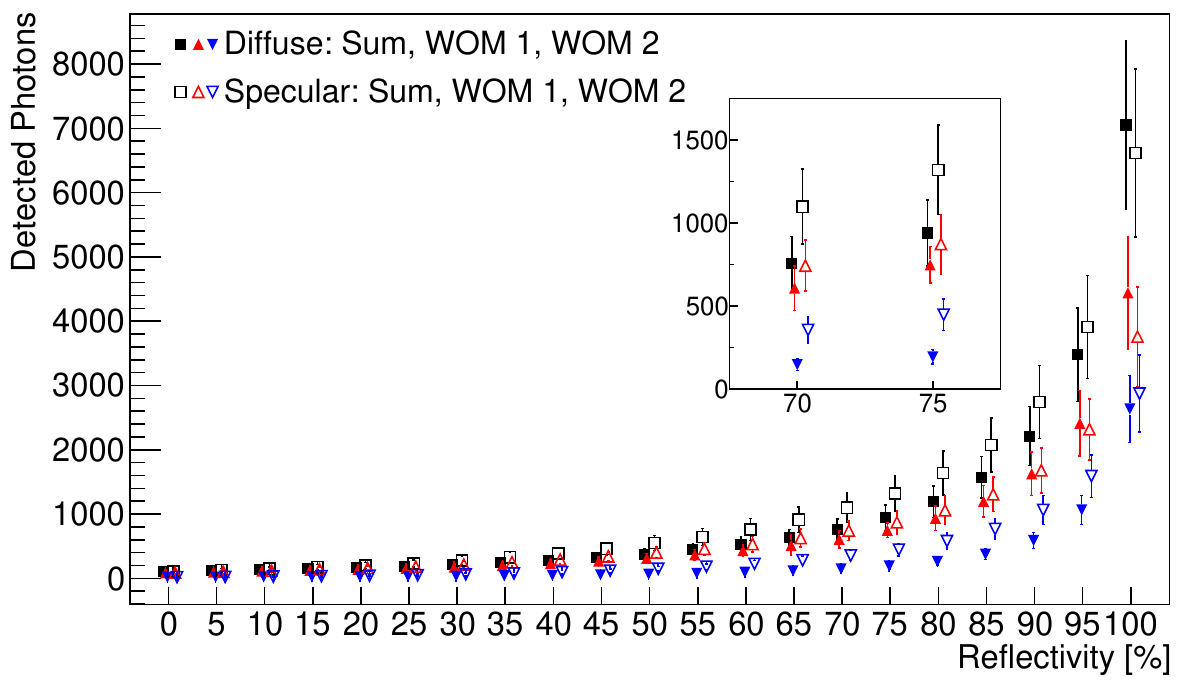}
    \caption{Simulation of the number of detected photons in the two WOMs and their sum as function of reflectivity for a purely diffuse and a purely specular reflector when a muon (MIP) crosses the detector cell in a corner.}\label{fig:SimulatedLightYield_Corner}
\end{figure}

For a muon crossing the detector cell in a corner, Fig.~\ref{fig:SimulatedLightYield_Corner} shows the number of photons that are detected by the individual WOMs as well as their sum, as a function of reflectivity for both purely diffuse and purely specular reflecting cell surfaces. For the same values of reflectivity up to 95\%, the specular reflector provides a higher sum signal than the diffuse reflector, while for reflectivities above 95\%, the opposite becomes the case. This can be explained by the fact that, for very large values of reflectivity, a photon can be reflected many times, resulting in path lengths up to 180\,m. According to the simulation, the average photon path length before hitting a WOM will be smaller for the diffuse reflector than for the specular reflector. Since the LS has finite transparency with an attenuation length of $\sim$8\,m in the relevant wavelength range (see Fig.~\ref{fig:tb2024_ls_absorption}), long path lengths will be suppressed, resulting in larger light loss for a high-performance specular reflector. Independent from the value of reflectivity, the difference in collected light yield between the two WOMs is always found to be smaller for the specular reflector, thus providing a more uniform detector response than the diffuse reflector.
Given the results of the measurements shown in Fig.~\ref{fig:reflectivity}, a reflectivity of 70--75\% is of particular interest for the materials explored here. In Fig.~\ref{fig:SimulatedLightYield_Corner}, this region is therefore zoomed in. 

\subsection{Light Collection}\label{Sec:CollectedLightSignal}

The amount of detected light measured in each event - the integrated yield - is quantified by integrating the waveforms of all SiPM channels over time, in units of V$\times$ns. The integration time of each waveform is taken to be 120\,ns, 20\,ns before the peak maximum and 100\,ns after the peak, in order to contain as much of the waveform as possible while minimising the effect of dark counts, crosstalk, and afterpulsing. The baseline of the waveform is corrected using the minimum root mean square slope of the waveform within a sliding window of 25\,$\times$\,0.3125\,ns width inside a 50\,ns time interval located well before the start of the signal. For more details, see Ref.~\cite{Alt:2023vuu}.
 
\begin{figure}[ht]
    \begin{subfigure}[c]{0.49\textwidth}
        \centering
        \includegraphics[width=1.0\textwidth]{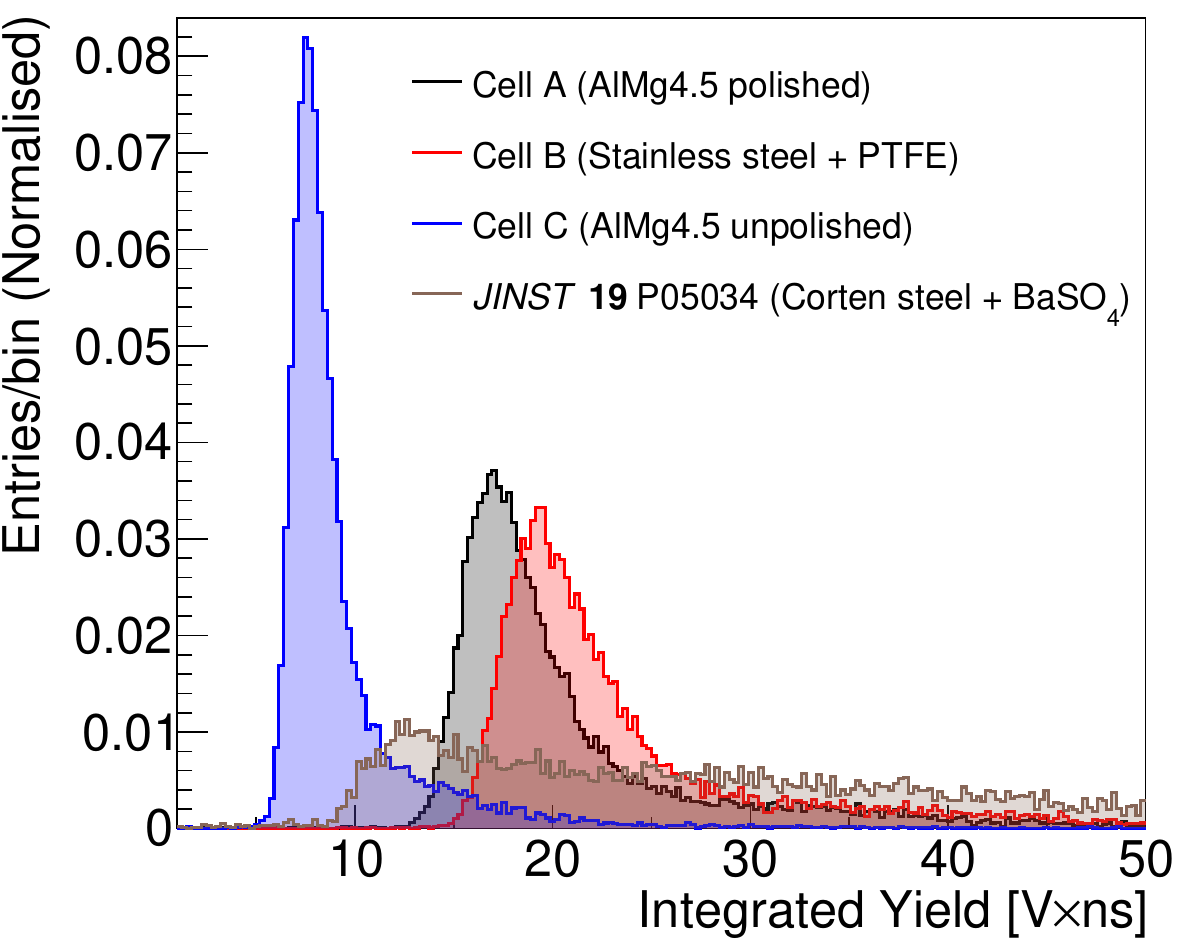}
        \caption{Beam crossing the cell at the centre.}\label{fig:charge-spectra_centre}
    \end{subfigure}
    \hspace{0.1cm}
    \begin{subfigure}[c]{0.49\textwidth}
        \centering
        \includegraphics[width=1.0\textwidth]{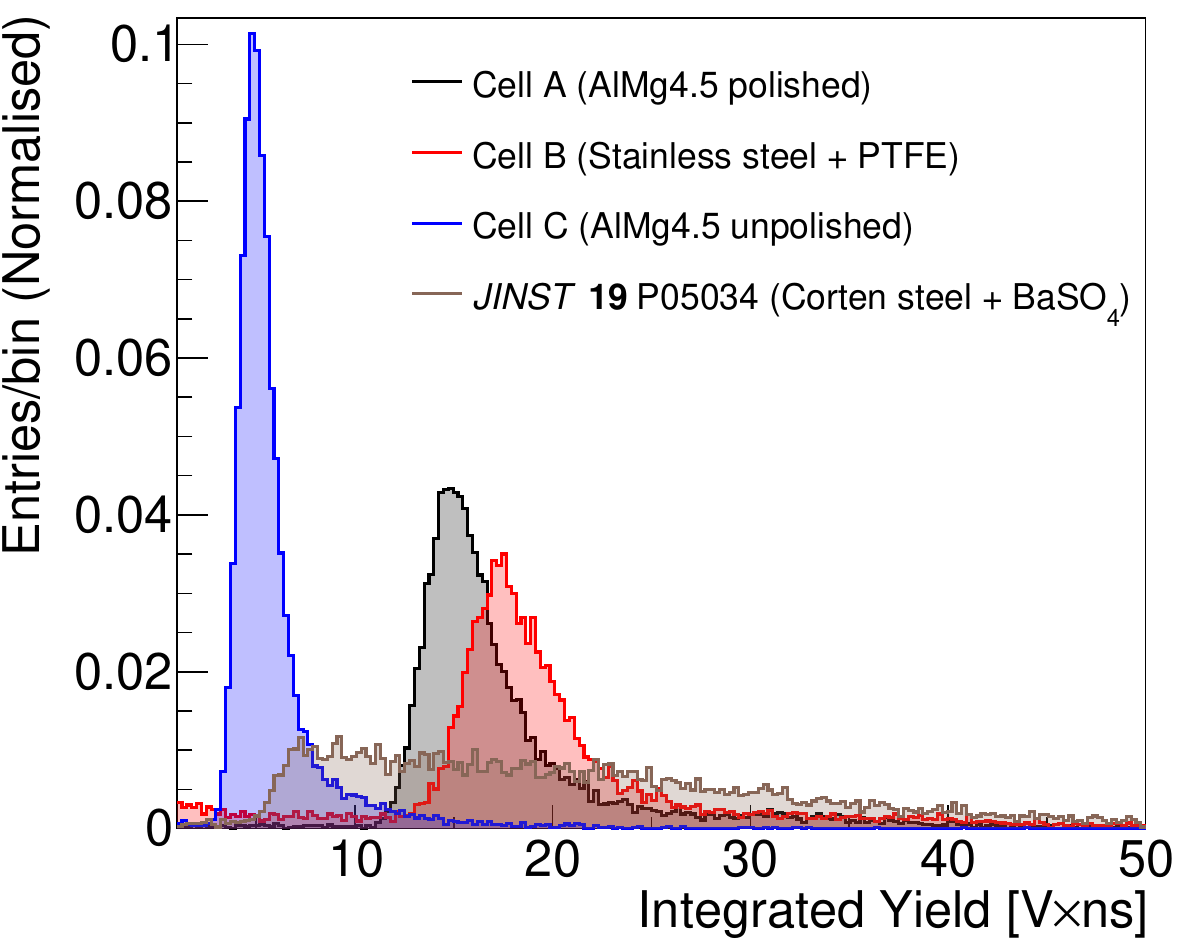}
        \caption{Beam crossing the cell at the corner.}\label{fig:charge-spectra_corner}
    \end{subfigure}
    \caption{Distribution of integrated yields, summed over all SiPMs  with the muon beam crossing the centre (\textit{left}) or the corner (\textit{right}) of the respective cell. Overlaid for comparison is the corresponding distribution observed for the previous prototype tested in 2022 with high-energy positrons at DESY~\cite{Alt:2023vuu}.}
   \label{fig:charge-spectra}
\end{figure}

For the three detector cells, Fig.~\ref{fig:charge-spectra_centre} shows the distribution of the integrated yield  summed over all 80~SiPMs in both WOMs when the muon beam hits the centre of each cell. For a comparison, the corresponding distribution is shown for the BaSO$_4$-coated Corten steel detector, which was tested with high-energy positrons at DESY~\cite{Alt:2023vuu}. The lower end of the energy deposition of the positrons correspond to the energy deposition of a MIP, and hence can be directly compared with the spectra measured in Cells~A, B, and C at the CERN PS muon beam. The highest integrated yield was achieved with Cell~B with the PTFE reflector. A slightly lower integrated yield was achieved with Cell~A (polished AlMg4.5). As expected, both Cell~A and B achieved a significantly higher integrated yield than the unpolished Cell~C, but also when comparing to the BaSO$_4$-coated Corten steel prototype, which is a very promising result. Similar results are achieved when the beam hits the corner of the cells (see Fig.~\ref{fig:charge-spectra_corner}). Particularly promising is that the MIP peak obtained with Cells~A and B is well above the dark count threshold of about 0.8\,V$\times$ns and that for both cells the relative difference in the integrated yield between the corner and the centre position of the beam is much smaller compared to Cell~C and also compared to the BaSO$_4$-coated Corten steel prototype. As a result, one can expect for a SBT detector built from cells of either type~A or B a good uniformity in detector response.

\begin{figure}[ht]
    \centering
    \includegraphics[width=0.9\textwidth]{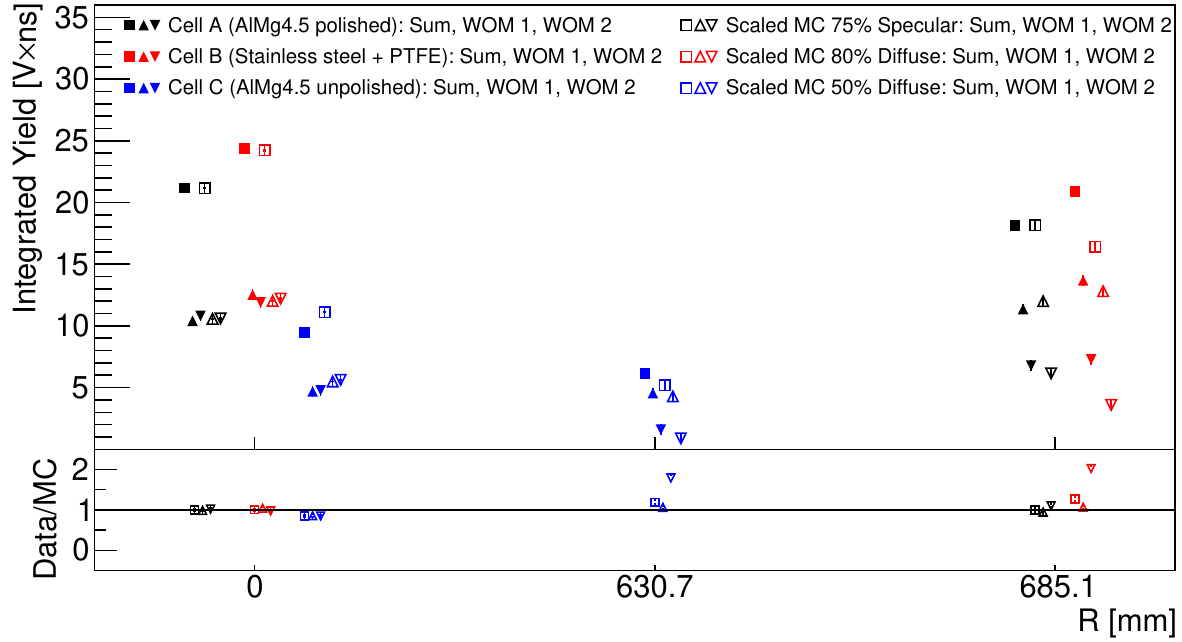}
    \caption{Mean value of the integrated yield for the individual WOMs as well as their sum for each of the three prototype cells (A, B, C), with the beam crossing the cell in the centre ($R=0$\,mm) or the corner (at a distance from the centre of $R=685.1$\,mm for Cells~A and B, and $R=630.7$\,mm for Cell~C, respectively). Solid markers represent measurement data, empty markers the MC simulation of the cells. The number of detected photons in the simulation is scaled to the values of the integrated yield measured in the data, as described in the text. For visual purposes, points with the same $R$~value are shown at a slight horizontal offset.}
    \label{fig:charge-spectra_centre_corner}
\end{figure}

This finding is confirmed by comparing the integrated yield in the two WOMs separately (see Fig.~\ref{fig:charge-spectra_centre_corner}), which shows the measured average values for the three cells in the centre and a corner of each cell. The measured results are compared to the expectation from the simulation for each cell, with Cell~A chosen as a reference for the comparison. The measurement in Sec.~\ref{sec:MaterialSpecs} of approximately 75\% specular reflection for the polished AlMg4.5 was taken as a baseline. The simulation results with 75\% specular reflectivity were scaled by a constant, such that the summed number of detected photons in the centre of the detector is equal to the summed integrated yield measured in Cell~A at the same particle crossing point. This scaling results in a very good match between the simulated and measured results in the corner point, as seen in Fig.~\ref{fig:charge-spectra_centre_corner}. Using the same constant scaling several diffuse reflectivities from 0--100\% were compared to Cell~B and C. The best match for these detectors is defined as the scaled simulation which results in the lowest difference in summed signals between simulation and data in the centre and corner of the cell. This results in an estimated diffuse reflectivity of 80\% and 50\% for Cell~B and C, respectively, which is in reasonable agreement with the reflectivity measurements in the laboratory.

\subsection{Detector Response Uniformity}\label{Sec:LightCollectionandDetectorResponse}

\begin{figure}[ht]
    \centering
\includegraphics[width=0.9\textwidth]{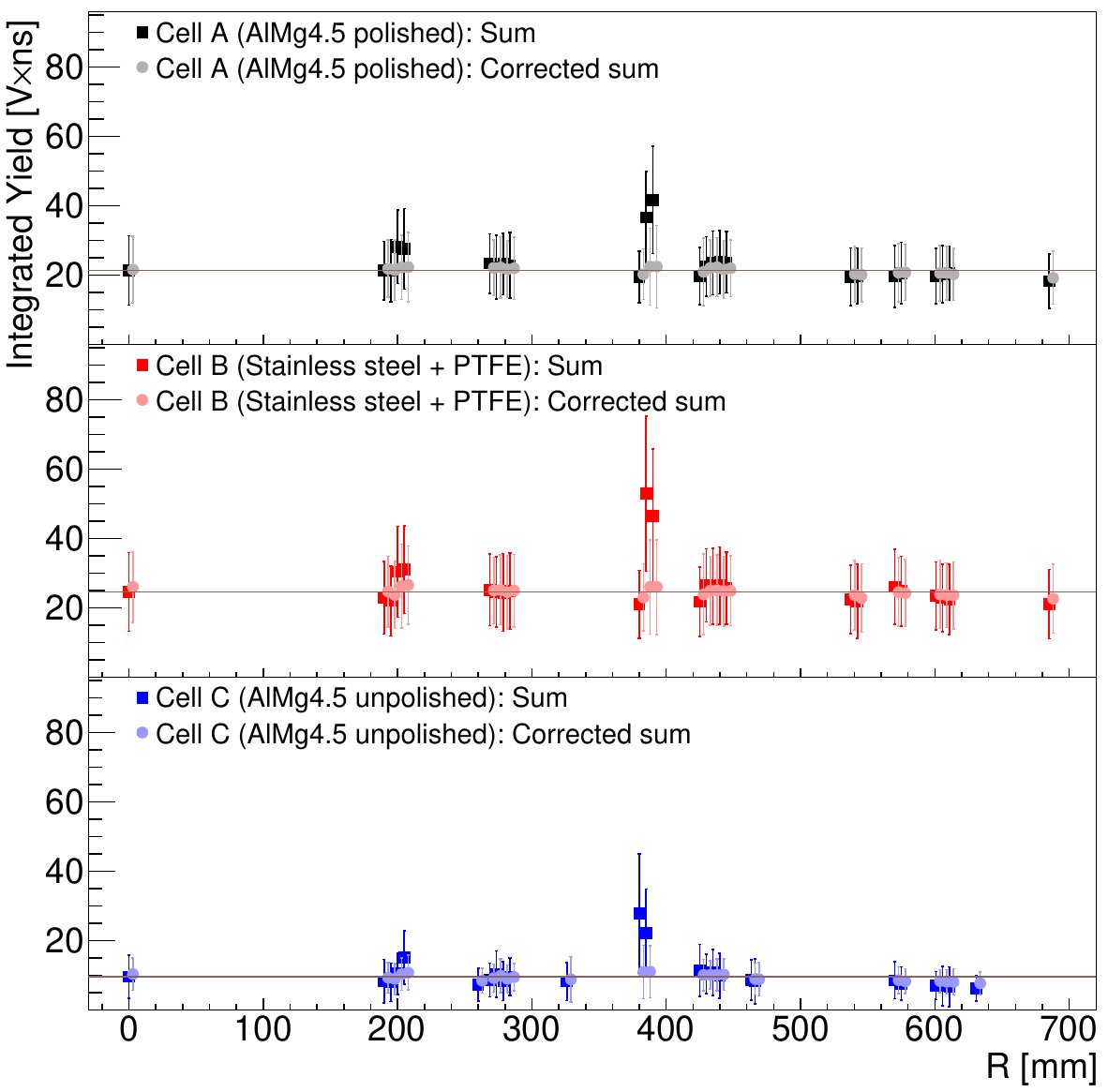}
    \caption{Mean integrated yield as a function of distance~$R$ between the particle crossing point and the centre of the cell for the respective prototype: Cell~A (\textit{top}), Cell~B (\textit{middle}), Cell~C (\textit{bottom}). The error bars are given by the standard deviation of the integrated yield distribution. Dark squares represent the measured values, light circles the same measurements after a likelihood-based correction (see text), with the horizontal line indicating the respective reference value. For visual purposes, corrected values and points with the same $R$~value are shown at a slight horizontal~offset.}
   \label{fig:chargepositions}
\end{figure}

The comparison between integrated yields for different particle crossing points is shown for all three cells in Fig.~\ref{fig:chargepositions}. Depending on the particle crossing point, the measured integrated yield can vary significantly.
As described in detail in~\cite{Alt:2023vuu}, it is possible to correct the measured integrated yield using a likelihood method that estimates the particle crossing point from the following measured quantities: the ratio of integrated yield between the two WOMs, the difference in signal arrival time between the two WOMs, and the detected integrated yield distribution over the eight SiPM groups of each WOM. After having applied the likelihood-based correction, the corrected integrated yield varies only by at most 5\% and 8\% between the different particle crossing points for Cell~A and B, respectively, as shown in Fig.~\ref{fig:chargepositions} by the light circles. For Cell~C, the likelihood-based correction also improves the detector response uniformity, however, here the non-uniformity goes up to 20\%.
 
\subsection{Time Response}\label{Sec:TimeResolution}

The time response of the LS-SBT prototype cells for a selected set of particle crossing points on the detector as sketched in Fig.~\ref{fig:deglocations} was examined. The signal arrival time for both WOMs was estimated using a constant fraction discrimination (CFD) method with a threshold value of 25\% of the maximum amplitude of each signal. For each event, the waveforms measured in the eight channels of each WOM were summed to reduce fluctuations from electronic noise. Afterwards, a smoothing algorithm was applied and the time at which the summed signal reaches 25\% of its maximum was determined by interpolation. From the resulting times, $\bar{T}_{1}$ and $\bar{T}_{2}$, the arrival time of the trigger signal, $T_\text{trigger}$, was subtracted, also using a CFD threshold of 25\%. Any variation in the time offset~$T_0$ from event to event cancels in this difference: $\bar{T}_{1, 2}^\text{corr} = \bar{T}_{1, 2}-T_\text{trigger}$. The uncertainty in the trigger time has been determined in earlier measurements to be approximately 0.2\,ns, which is significantly smaller than the achieved time resolution of the WOM signals.

\begin{figure}[ht]
    \begin{subfigure}[c]{0.49\textwidth}
        \centering
        \includegraphics[width=1.0\textwidth]{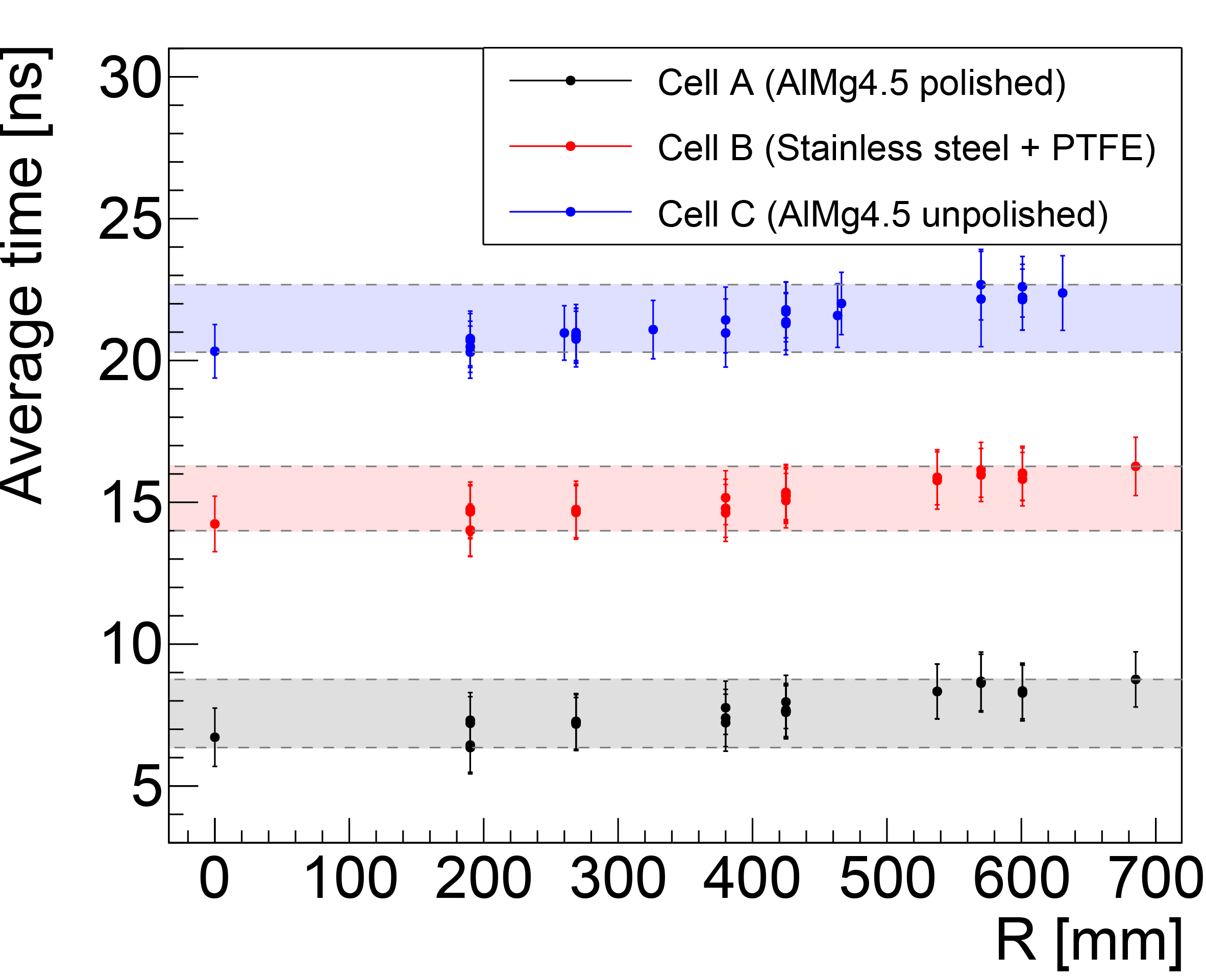}
        \caption{Before correction.}\label{fig:time_tot}
    \end{subfigure}
    \hspace{0.1cm}
    \begin{subfigure}[c]{0.49\textwidth}
        \centering
        \includegraphics[width=1.0\textwidth]{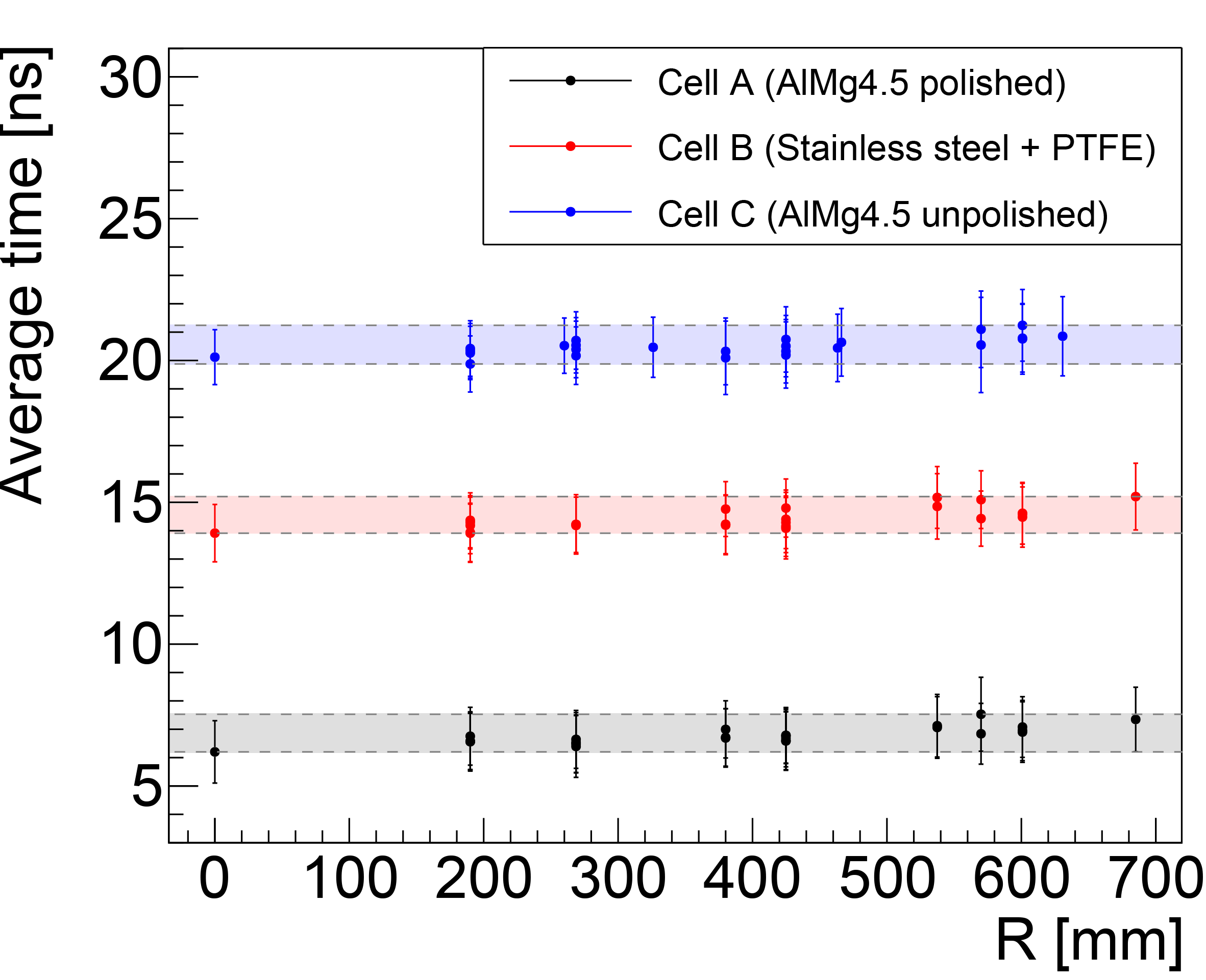}
        \caption{After correction.}\label{fig:likelihood_timeavg}
    \end{subfigure}
    \caption{Estimated particle arrival time $\bar{T}_{12}$ at the detector, as a function of distance between the particle crossing point and the centre of the cell for the three prototypes. Before (\textit{left}) and after~(\textit{right}) application of the likelihood-based correction. The measured fractions of integrated yield in each WOM and in each SiPM group of each WOM, as well as the difference in signal arrival time between the two WOMs, are used to reconstruct the most likely particle crossing point for the respective event. The error bars represent the standard deviations of the distributions, the shaded bands depict the time variation over the entire box. The offset in average time between the three cells is arbitrary.}
\end{figure}

For each particle crossing point, the particle arrival time at the detector cell was estimated for each event by calculating the average between the two times: $\bar{T}_{12} =(\bar{T}_{1}^\text{corr} + \bar{T}_{2}^\text{corr})/2$. The mean and the standard deviation of $\bar{T}_{12}$ for different particle crossing points are shown in Fig.~\ref{fig:time_tot} for the three different cells. Across all particle crossing points tested, $\bar{T}_{12}$ varies within a 2.1\,ns time window. Hence, the systematic uncertainty on the time response from the physical size of the detector cell is about $\pm{1}$\,ns. As expected, the standard deviations are larger for Cell~C compared to Cell~A and B, since the number of detected photons in Cell~C is smaller than in the other two detectors. The standard deviation of the $\bar{T}_{12}$ distributions in Cell~A and B is about 1.0\,ns. As a result, we do not correct the standard deviation of the $\bar{T}_{12}$ distributions for the uncertainty on $\bar{T}_\text{trigger}$.

Without further correction, the intrinsic time resolution of the detector is $\pm{1}$\,ns, due to the time variation in $\bar{T}_{12}$ if the particle crossing point is unknown, plus the statistical uncertainty as reflected by the standard deviation. To reduce the intrinsic time variation, the same likelihood-based procedure as described in Sec.~\ref{Sec:LightCollectionandDetectorResponse} to reconstruct the particle crossing point and correct the average time is applied. The results of the correction are shown in Fig.~\ref{fig:likelihood_timeavg}: the intrinsic (systematic) time resolution is significantly reduced to a time response variation within a 1.2\,ns ($\pm{0.6}$\,ns) time window, compared to the time window 2.1\,ns ($\pm{1}$\,ns) before the likelihood-based correction. 

\section{Summary and Outlook}\label{Sec:Summary}

Three full-scale prototype detector cells for the SHiP Surrounding Background Tagger (SBT) were constructed of the aluminum-magnesium alloy AlMg4.5 and stainless steel, respectively. Both materials are known to provide good chemical compatibility with the liquid scintillator made from LAB~+~PPO foreseen for the SBT. To reach a high reflectivity at the relevant wavelength around 380\,nm, the inner walls of the stainless steel cell were clad with 0.5\,mm thick PTFE sheets, while the inner walls of one of the two AlMg4.5 prototypes were manually polished. A third cell made from AlMg4.5 without any further surface treatment was studied for comparison. Each liquid scintillator-filled cell was equipped with two Wavelength-shifting Optical Modules collecting the primary scintillation photons and guiding the secondary photons to an array of 40\,SiPMs. These were read out in eight groups of five SiPMs. The detector prototypes were tested  with 5\,GeV muons in November 2024 at CERN's PS T10 testbeam facility, with the goal of directly comparing their performance in detected light yield.

We find that the stainless steel cell clad with PTFE sheets produced the highest detected light yield. In the polished AlMg4.5 cell, the integrated yield was found to be 15\% lower. For a beam position on a corner of the cells, the mean integrated yield detected in the WOM further away from the particle crossing point was found to be a factor of 9.1 (8.5) above the dark count threshold for the stainless steel~+~PTFE (polished AlMg4.5) cells, respectively. Using a likelihood-based method, a detector response of integrated yield that is uniform within~8\% can be achieved. With the same method, we reach a variation of the time response of $\pm 0.6$ ns, with a time resolution of $1.0$ ns. 

For both the polished AlMg4.5 cell and the PTFE-clad stainless steel prototype, the achieved detected light yield and time resolution meet the requirements of the SHiP Surrounding Background Tagger. Concerning another crucial parameter, the uniformity of the detector response across the detector cell, the polished AlMg4.5 cell shows the best performance among the tested prototypes. The results presented here will significantly affect the final design of the SHiP Surrounding Background Tagger.

\section*{Acknowledgements}
The measurements leading to these results have been performed at the CERN East Area test beam facility. We especially thank the CERN test beam coordinators for their constant support, as well as the safety and transport groups. We are grateful for the skills and efforts of the technicians of our institutions. We thank Stefania Bordoni from the University of Geneva for providing the WaveCatcher crate that was used for the data acquisition. We thank Dominique Breton and Jihane Maalmi from IJCLab for their advice on running the WaveCatcher. We thank the Deutsche Forschungsgemeinschaft (DFG) for funding support within grant 289921825 and the Bundesministerium für Bildung und Forschung (BMBF) for funding support within the \hbox{High-D} consortium. The test beam measurements have received funding from the European Union's Horizon Europe Research and Innovation programme under Grant Agreement No.~101057511(EURO-LABS).

\end{document}